# USER PERCEPTION OF PRIVACY WITH UBIQUITOUS DEVICES

Priyam Rajkhowa

Indian Institute of Science, Bengaluru priyamrajkhowa56@gmail.com

Pradipta Biswas

Institute of Science, Bengaluru, pradipta.iisc.ac.in

Privacy is important for all individuals in everyday life. With emerging technologies, smartphones with AR, various social networking applications and artificial intelligence driven modes of surveillance, they tend to intrude privacy. This study aimed to explore and discover various concerns related to perception of privacy in this era of ubiquitous technologies. It employed online survey questionnaire to study user perspectives of privacy. Purposive sampling was used to collect data from 60 participants. Inductive thematic analysis was used to analyze data. Our study discovered key themes like attitude towards privacy in public and private spaces, privacy awareness, consent seeking, dilemmas/confusions related to various technologies, impact of attitude and beliefs on individuals' actions regarding how to protect oneself from invasion of privacy in both public and private spaces. These themes interacted amongst themselves and influenced formation of various actions. They were like core principles that molded actions that prevented invasion of privacy for both participant and bystander. Findings of this study would be helpful to improve privacy and personalization of various emerging technologies. This study contributes to privacy by design and positive design by considering psychological needs of users. This is suggestive that the findings can be applied in the areas of experience design, positive technologies, social computing and behavioral interventions.

**Keywords**

## 1 INTRODUCTION

Camera and photographic technology were well documented by Han Chinese and Arab physicists in mediaeval period. The first commercial digital camera was available in 1975 [7]. Today with abundance of smartphone and in-built high end digital cameras, new technical possibilities and social habits have emerged. A popular proverb says that a photograph can say a thousand words. People often take pictures of themselves, known people, objects and makes them accessible to public through social networking websites. New technologies like augmented and mixed reality systems use live video feed from ubiquitous devices and serve a plethora of purposes like navigation, inspection, localization and so on. However, this abundance of cameras and photographs also create serious challenges for both user and bystanders' privacy [50]

Privacy is a state in which one is not observed and disturbed by other people. It is a state of being free from public attention. It is a fundamental right which is important to determine autonomy and to protect the dignity of a human being so that it serves as the foundation or base upon which several other human rights are built [50]. Privacy plays an integral role in every human being's day to day life. Communication privacy management theory (CPM) investigates the concept of privacy from multiple ways and explains privacy taking into consideration various contexts.

Communication privacy management theory (CPM) was first introduced by Sandra Pedronio in 2002. It is an evidence-based theory which provides explanations of how various decisions are made in order to disclose and protect private information. CPM focuses on the relationship that people have with each other in communicative text like face-to-face interactions, on social media, and in dyads or groups [44]. This theory is based on a communicative social behavioral perspective and argues that because people consider private information something that they own, over which they desire control, individuals feel violated when others find out something about them without their permission. CPM contends that boundaries are multiple in nature, humans regulate ownership and concurrent control for both personal and collective boundaries. [44]

A qualitative study was undertaken to understand perception of privacy in a comprehensive manner. There are not many studies done that captures user's perspective of privacy invasion specifically with regard to public and private spaces. This study aimed to represent various attitudes and beliefs held by participants in both these spaces. It would help to influence negative attitude like considering surveillance as a threat and that privacy is a myth in digital era. Moreover, it would facilitate designing ubiquitous devices in a way that would incorporate psychological well-being. This study also explored behavioral and attitudinal components of privacy perception in public and private spaces. Findings would contribute to designing and production of enhanced privacy products, developing socio technical solutions for privacy risks, facilitate adoption of technologies, promote sensitivity towards privacy, design behavioral interventions and persuasive technologies. The upcoming section would throw light on literature.

## 2 REVIEW OF LITERATURE

Privacy considerations with emerging ubiquitous technologies is not a new concern. Concerns related to privacy of wearables, camera, telecommuting, live streaming are being explored [63,66,6,16]. Understanding the dynamic nature of privacy along with its implications is a bigger challenge. In this study we reviewed research papers, articles, webpages related to meaning and functions of privacy, theories of privacy, concerns related to privacy invasion, users and bystanders' privacy concerns and lastly technical solutions available to address problems of privacy invasion.

Privacy is like treasure that all wants to protect. It unlocks aspects of an individual that are of utmost importance, intimate and personal that could make anyone vulnerable. Over the past centuries, the meaning and definition of privacy has dynamically changed and many privacy theorists agreed that it comprised of multiple dimensions. Privacy has evolved from the notion of "the right to be alone" [9] to "the condition of not having undocumented personal knowledge about one possessed by others." [53]. There are multiple theories of privacy such as privacy regulation theory, non-intrusion theory of privacy [48], Alan Westin's theory of privacy [47]. Privacy is necessary for human well-being and flourishing [49]. It also serves certain functions like personal autonomy, emotional release, self-evaluation and protected communication [55].

There are risks associated when needs of privacy are not met. Privacy problems may not always conform with the definitions of privacy. They could range from adversities related to information collection, information processing, information dissemination and invasion [67] Invasions could be of various kinds like extraction, observation and intrusion. [40]. With emerging technologies, it was important to address the different kinds of privacy and also find solutions for the same. Finn et al. [24] described seven different types of privacy, namely- privacy of the person, privacy of behaviour and action, privacy of personal communication, privacy of data and image, privacy of thoughts and feelings, privacy of location and space and privacy of association (including group privacy).

A rising concern for general public was escalating and abundant availability of smartphone cameras. A bystander is a person who is present but not taking part in a situation or event and is a chance spectator. Bystanders' privacy arises when a device that collected sensor data could be used to identify third-parties (or their actions) when they had not given consent to be part of the collection. [57, 58, 59]. Motti et al. [50] undertook a study that found the following concerns related to bystander's privacy of wearables: facial recognition, social implications, social media, user's fears: surveillance and sousveillance, speech disclosure, surreptitious A/V recording and location disclosure.

There are multiple technological solutions available for privacy related problems. These solutions differ in terms of effectiveness, usability and power consumption. Various technological solutions available are sensor saturation, disabling messages via data communication interfaces, context-based interfaces [34, 69] to solutions that hide the identity of bystanders to avoid their identification. [26,41, 65, 34]. I-Pic [2] was a software platform that integrated digital capture with user- defined privacy. The user chose a level of privacy like if image capture was allowed or not based upon social context such as public, with friends, workplace. Moreover, Raval et al. [64] proposed a solution that required physical tagging of objects and locations. On the other hand, Jung et al. [37] used explicit user actions like gestures which conveyed the privacy choices. There are other methods [80] that combined part-based models and deep learning by training pose-normalized Convolutional Neural Networks (CNN). It inferred human attributes like gender, hairstyle, clothing style, expression, action from images of people under variation of viewpoint, pose, appearance, articulation and occlusion. In addition, FacePet was a pair of intelligent goggles which used artificial intelligence (AI) algorithm that prevented unauthorized face detection operated via mobile application that does not rely on third-party or systems. [58]. There was another solution that used computer vision for locating all retro-reflective CCD or CMOS camera sensors in a protected area. HideMe was a framework that preserved the associated users' privacy for online photo sharing [42]

Hatuka [29] conducted an experiment on smartphone users that included interviews and the use of a smartphone android application that combined online tracking with experience sampling. The findings reported that there was a convergence between the online and offline worlds (a 'public' situation in the offline world is also considered as such in the online world) which was a condition that contributed to the normalisation of 'asymmetrical visibility'. This possibility assumed that "individuals are actively aware agents who understand that contemporary visibility practices provide them with tools to control their social interactions, such as setting meetings and messaging as well as 'crafting' their appearance in 'public', but at the same time limit their privacy" [24]

This study aimed to understand life experiences, beliefs and perception regarding various concerns related to protection of privacy. It also explored and understood various actions that people took to protect themselves from invasion of privacy. It examined why participants choose certain methods of protecting themselves from invasion of privacy. This survey investigated various actions taken by individuals to prevent privacy invasion. It also explored various divergent ways of how people protected their privacy. Open or inductive survey was used in this study to empirically search for diversity of attitude, beliefs and behavior of participants. It provided with an opportunity to recognize various categories and combinations of privacy concerns which existed in the mind of the sample group.

### 2.1 Ethical Protocols

Institutional approval was obtained prior to approaching the sample for the research study. Informed consent was obtained from all participants prior to commencement of responding to the questionnaires. They were also informed of their right to withdraw from the study if they wished to do so at any given point of time along the duration of the study. Privacy and confidentiality of the participants was ensured. It is guaranteed that the participants faced no physical or psychological harm as a result of their participation in the study.

## 3 METHODOLOGY

In the upcoming subsections we shall discuss the various methods that was used to conduct this research. It would explain the type of research, method of data collection, materials used for data collection, analysis of data and the rationale for choosing these methods.

### 3.1 Research questions

Research questions are systematic inquiry that facilitates collecting data from participants. The following research questions helped us to gather information pertaining to participant's experiences and perceptions in line with the objectives of this study.

- What are the various attitudes and beliefs that contribute to user perception of privacy invasion?
- What are the experiences of participants with invasion of privacy with context to spaces- public and private?

- What are the various thoughts held on privacy risks held by people in public and private spaces?
- What are the actions taken to protect oneself from privacy invasion in public and private spaces?

### 3.2 Paradigm

Phenomenology approach is an approach used to understand the nature of experience of the participants. [18] It emphasises on core experiences and perspectives of participants. It also interprets the understanding of their experience. The elementary goal of this approach is to arrive at a description of the nature of phenomenon. It tries to cater to what did participants experience in terms of the phenomenon and what context influenced these experiences. [18]

### 3.3 Objectives

The study aimed to understand experiences and discover attitude of individuals related to privacy invasion. It uncovered viewpoints regarding invasion of privacy and gained an understanding between attitude and behaviour towards privacy invasion. It also sought to uncover if any attitude behaviour gap existed. Objective of this research was to understand:
- Attitude of participants regarding invasion of privacy.
- Experiences of participants related to privacy invasion.
- Relationship between attitude and behaviour of privacy invasion.
- Attitude behaviour gap of privacy invasion.

### 3.4 Sampling procedure

Purposive sampling was used as it had the advantage of familiarity, comfort, access, easy mode of communication was some of the important considerations around purposive sampling. Mid-range sample size (60-99) was chosen to ensure dataset richness address research questions [10,11, 23, 25].

### 3.5 Participants

A total of 60 participants (24 female, 36 male) undertook this survey. 70% participants belonged to the 25 to 34 years age group category. 10% participants belonged to 35 to 44 age group categories. Figure 1 is a diagram that represented the age, countries of residence and mother tongue of the participants.

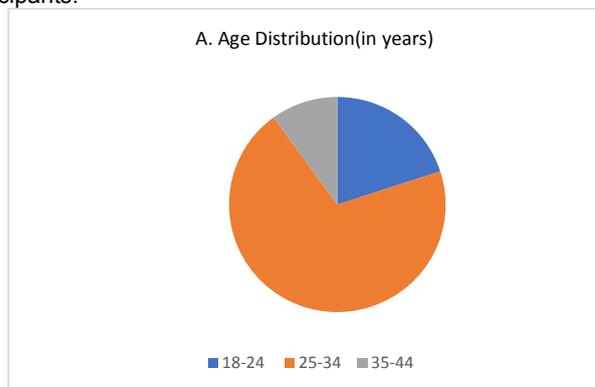

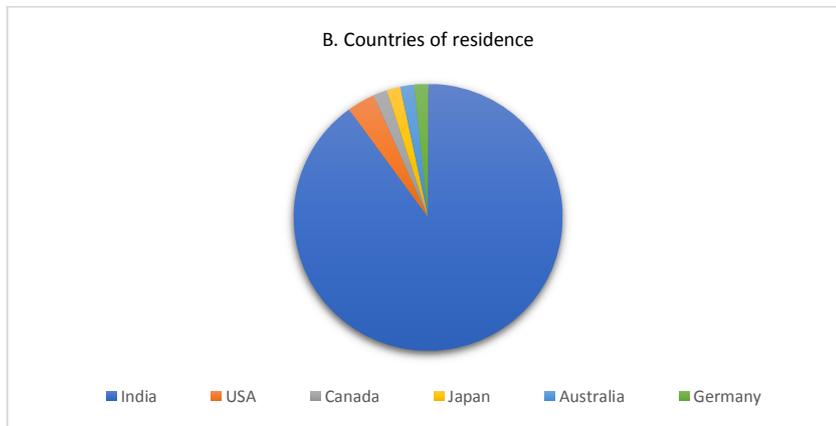

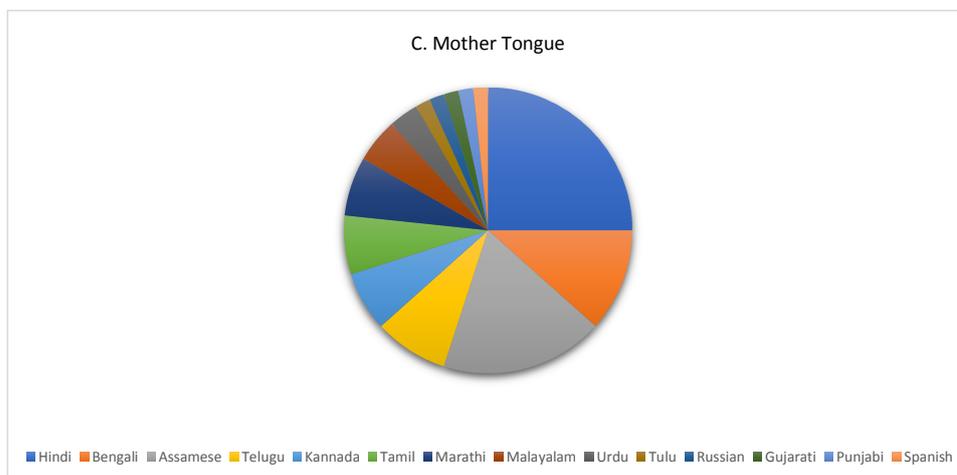

Figure 1: Sample distribution of participants.

Participants were from four different countries- India, Pakistan, Kazakhstan and Venezuela. 95% of participants were from India belonging to states: Karnataka, West Bengal, Assam, Tamil Nadu and Maharashtra. 75% of participants completed post-graduation while 23.33% of participants were undergraduates. Participants were working professionals in the area of service, healthcare, marketing and law.

Only one participant belonged to Hispanic ethnicity. All other participants belonged to Asian ethnic community. All could read and write in English language. Mother tongues of participants were Hindi, Bengali, Assamese, Kannada, Gujarati, Spanish, Russian languages.

Informed consent was obtained and confidentiality of the responses and identity was ensured. All participants were encouraged to respond honestly and patiently.

**4 PROCEDURES**

Qualitative survey was used as it ensured maximum heterogeneity of samples which focused on diversity and not on typicality, it was affordable and large geographically dispersed populations was accessed, it encouraged disclosure of sensitive topics like revealing information of what respondents does in their private space to prevent invasion of privacy, non-direct contact of the participants and researcher enhanced personal safety of both parties. It did not burden participants who did not have access to great internet connectivity. A survey questionnaire in English language was drafted. It consisted of questions that were open, short, clearly explained, avoided ambiguity and assumptions. It incorporated the diverse experiences and views of the respondents. It comprised of demographic, closed and open-ended questions. [10,11]

Standard demographic clicks and checked box closed response questions with the option of 'other- please specify' option was used. Closed ended question was followed with a clarificatory secondary question like 'please explain'. Open ended questions provided the participants to define their thoughts and ideas. These open-ended questions were framed in a broad manner. To ensure flow, the questions on similar aspects of a topic were clustered together. The survey aimed to capture respondents' lived experiences record. Ten questions on lived experiences, perceptions and actions were asked to ensure better focused responses and to prevent participant disengagement/ fatigue. Participants were encouraged to spend time and reflect on their responses. [11, 14, 23, 31]

The questionnaire was evaluated by two experts and was piloted on eight participants. Recommended suggestions and insights from pilot study were incorporated in the final survey questionnaire.

### 4.1 Data collection
Data was collected at one shot due to prior knowledge of the context and constraints of time. Google forms was sent out early of March 2021 via e-mail. The questionnaire was also forwarded to campus email groups of Indian Institute of Science, Bengaluru. Data was collected till the end of March 2021. The response rate of the survey was 93.75%.

### 4.2 Data analysis
Inductive thematic analysis was used to analyze data. This meant that analysis was not shaped by any existing theory, it was drawn from the collected data. Data was systematically summarized and transformed from vast quantity of text into a highly structured and refined result. It was sorted into numerous manageable categories on a variety of levels like word phrase, sentence/theme. It guided to form meanings and relationships of the words and concepts. [11,13]

Familiarizing oneself with the data was the first step. The responses were collectively organized with respect to each question in Excel sheet. All answers were carefully read through multiple times. Firstly, each individual survey questionnaire was read through thoroughly. Secondly, each group of answers pertaining to sole questions was read various times. This enabled to get a holistic meaning of responses. Notes were made after five to six thorough reading of all the responses. Initial impression was noted down. Objective was to understand the context of the sample population. [11]

The data was then divided into smaller parts simultaneously ensuring that the core is retained. This was done using coding units. A code is a label that most exactly describes what this particular condensed meaning is about. They are usually one or two words long. It makes identification of connections easier between various meaning units. [11, 18] The coding units were applied to sort data into categories. A category is formed by grouping together the codes that are related to each other through their content/ context. Commonalities and differences were looked for in the most objective and accurate way that the participants responded. Contradictions were identified and noted as well. The references, attitudes and characteristics of the responses was identified and marked. The themes were reviewed. Thematic map was created, provisional themes and subthemes were explored and relationships between them was inspected. Finally, the themes were defined, named and final analysis was written. [10, 11] The upcoming section would run through various themes and subthemes found from the online survey. It would highlight multiple contexts of participants and enlist verbatim of participants for better understanding and painting a meaningful picture of the patterns observed in data.

**5 Results**

The study aimed to explore and understand experiences related to privacy invasion. The context of participants is described in the first section and the later part of this section describes the themes and its sub themes. Discussion section further explores and investigates the themes and contextualises them with relevant literature. A short and crisp summary of insights is presented after the discussion that would be useful for practitioners.

### 5.1 Participant's context

This section on participant's context provided a comprehensive understanding of the participants like self-perception of oneself related to various factors that nourishes attitude formation of privacy. Factors like perception of how good one is with usage of technology, if they used ubiquitous devices like smartphones and laptops regularly, if they clicked pictures of oneself and others in public spaces or not, and so on provided us with an estimation if they would have experienced any sort of privacy invasion. Our study found that:

- 63.33% of the participants considered themselves to be good with the use of technology. It meant that participants knew methods, theory, and practices of various technology. 31.66% of participants considered themselves to be tech savvy. It meant that they considered themselves to be well learnt, informed about the modern technology and also consciously used their skills to merit current technology. It also meant that this section of population used smart devices with diverse software tools and applications in it.
- 71.66% of the population sample used smartphones on a daily basis. 21.66% participants used laptop. Other devices and gadgets used were computer and portable Bluetooth devices.
- Largest number of respondents (33.33 %) frequently took pictures and videos of oneself and others. 21.66 % of the respondents clicked pictures of themselves or others once in 15 days. Very few of them took pictures of themselves and others once a year to rarely.
- 91.66% of the respondents visited public places like shopping malls, restaurants, public transport, gyms and fitness centers, parks. It was noted that these public places were visited (40%) on a weekly basis, monthly (33.33%) and then daily by respondents. About 8.33% of the respondents visited public spaces once in 2 days. A few of the public places visited by a few respondents were clubs, court and teashops.
- 66.66% of the respondents admitted that they took pictures and videos in public spaces. 80% of the respondents reported that they very rarely attended video calls in public.
- About 48.33% of the respondents reported that they spotted strangers in pictures which was clicked by them in public spaces. This meant that they may have violated facial privacy of bystanders.
- Majority (21.66%) of the participants were comfortable being captured in photographs/videos by their friends. 8 out of 60 participants were comfortable being captured by a professional photographer. The next choice of people that participants were comfortable getting clicked were their colleagues, CCTV camera and media personal. Only 4 out of 60 participants were comfortable being captured in photographs by a stranger.

- 93.33% of people confirmed that they had respected the privacy preferences of public around them when they clicked pictures/videos.
- Participants responded that they would avoid taking pictures of both women and children in public spaces. On the other hand, 73.33 % of participants stated that they would completely avoid taking pictures of women in public. A few participants mentioned that they would avoid taking pictures of any strangers and people belonging to LGBTQ community in public spaces.
- Public spaces where participants would avoid taking pictures/videos were- restrooms, gyms, hospital, movie theatres, places of worship, shopping malls, parks, streets.
- Spaces considered as private by the participants: home, office space, seating table in restaurants, washrooms, private car, gym or fitness activity area, places of worship, theatres, trial rooms in shopping malls.
- Places like gyms, movie theatre, places of worship were considered both public and private.

Sources through which participants kept themselves updated about various privacy regulations were: curated list of pro- privacy groups on telegram, forums and subreddits dedicated to privacy. For example, the subreddit "r/DE google", blogs, news articles, twitter, google, social media. 30% of participants did not keep themselves updated about various privacy regulations.

**5.2 Themes and subthemes:**

The analysis of the online survey questionnaires revealed the following five themes: attitude towards privacy invasion, awareness about privacy (risks and regulations), seeking permission, dilemmas concerning $S^3$ (safety, surveillance and sousveillance) and actions preventing invasion of privacy. The first three themes broadly described thoughts, expectations, beliefs, and desires pertaining to attitude of privacy, the fourth theme recounted confusions associated towards privacy and the last theme explained actions or behavior undertaken by participants to maintain privacy. All these themes were common for both public and private spaces. Each of these themes constitute of diverse dimensions that was reported as sub-themes. We acknowledged that some of these themes and their underpinning dimensions may overlap to some degree. Verbatim of participants was reported under sub themes. They are described below:

5.2.1 Attitude towards privacy invasion

Attitude towards privacy invasion in public space: Attitude can be defined as a relatively enduring and general evaluation of an object, person, group, issue, or concept on a dimension ranging from negative to positive. [74]. This theme represented various beliefs and ideas held by participants towards privacy invasion in public space. As expressed by many participants respecting privacy of others in public spaces was considered as a moral responsibility. Disrespecting privacy made people feel uncomfortable and disturbed. Participants mentioned that they disliked when they were photographed by friends and strangers without their permission. Explicit consent was considered essential irrespective of the nature of interpersonal relationship one had with bystanders. Consensus of all participants was on the idea that privacy of others had to be respected as they expected others to do the same. It was expected that people-maintained decency in public spaces. Clear and explicit pre-defined rules should be put up on punishment when privacy was invaded.

*"It's basic manners that we be respectful of others around as 1. they might not want to be disturbed and 2. They could be uncomfortable with the thought of being included in the picture."*

*"I personally sometimes do not like if anyone takes a picture of me suddenly out of nowhere even if he or she is a friend, especially if I am in an uncomfortable situation. Hence, I keep the same thing in mind when it comes to other people too. Everyone has different levels of privacy and I respect that and I would never take a picture of someone without taking their consent."*

Some participants also mentioned that they despised when strangers photobombed in their pictures, hence they preferred to wait till strangers in the intended frame cleared out. There was a general consensus among participants that pictures taken in public were usually uploaded in social media, therefore it was further important to seek permission or consent. Also, as reported by participants, seeking permission was not always possible so, they resorted to informing bystanders that they were involuntarily going to become a part of a picture. The choice was left upon bystanders' if they wanted to be a part of the picture or recording. It was expected from bystanders (strangers and friends) to first inform that they were about to film/click picture. Participants also mentioned that they would delete pictures if anyone did not want themselves to be captured or their pictures be uploaded in social media.

*"Permission should be asked always if we want to click a picture and also if any other person is not willing to appear on the picture. As often we click the picture we tend to post in social media these days. Hence, it matters whether we respect the privacy or seek permission from the other people"*

*"If someone doesn't want to be tagged on social media then that choice must be respected for we are in a generation that clicks pictures only to be posted on the internet"*

*"I would politely inform others (known or unknown) that they are involuntarily becoming a part of a picture that I'm taking. Whether to be in it or not, is upto them."*

*"I don't like it when a person gets in the way of my picture so I would rather not take a picture if someone's in the way."*

Fifty-four participants agreed to respecting others privacy in public space. Seeking permission was considered important as participant thought that every individual had their exclusive notion of boundaries and it should be respected. Participants believed that not all people in public spaces were aware if they were violating privacy of others. So, a few participants considered that due to lack of awareness, violation of bystander's privacy could be a mistake that was not committed on purpose. 8.33% of the participants expected that people in public space should not look over bystander's phone screen. They also had negative attitude towards eavesdropping and starring in public space. In shopping complexes, CCTV facing trial rooms were unacceptable. If groups were keeping to themselves then their privacy should not be invaded.

*"It is very important to understand boundary and space. People should be respectful towards others' privacy and their preferences of being photographed or not."*

*"I don't think people do it on purpose, theres just a lack of knowledge etc. As for protecting myself, don't have many expectations since nobody has run up with a camera to photograph me etc"*

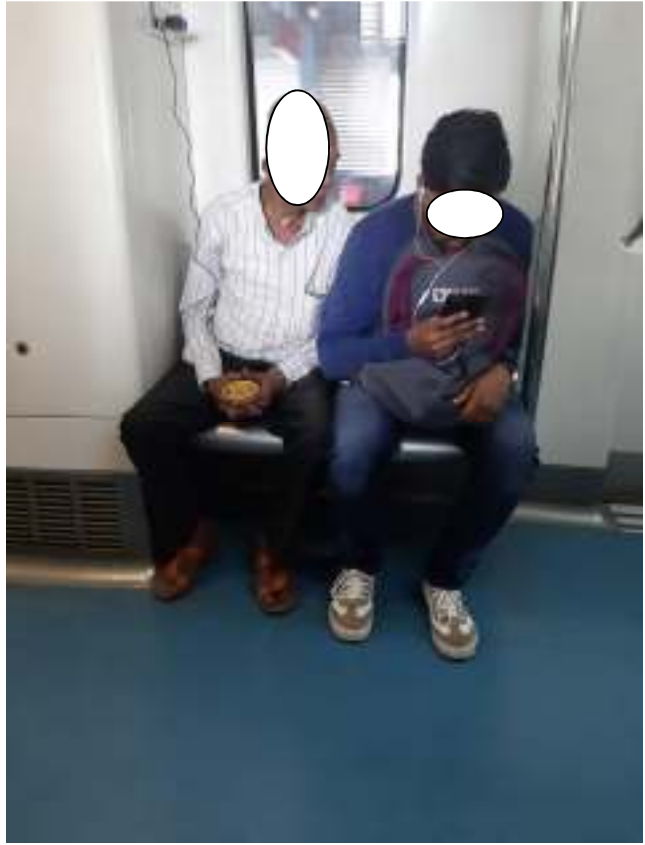

Figure 2: Image depicting multiple types of privacy invasion in public space.

Two participants mentioned that they would not consider respecting privacy of people in public space if bystanders were in the process of committing a crime/ wrongdoing. Three participants considered clicking selfies as a private act. When bystanders deliberately captured themselves into frame of selfie, it was considered as a violation of the participant's personal privacy. Four participants thought that expecting complete privacy in public space was unreasonable and chaotic.

*"Expecting complete privacy in public spaces is unreasonable, as the space is not meant to be private."*

*"The bubble of private space around you must be respected whether it is private or public"*

Attitude towards privacy invasion in private spaces: Participants preferred to avoid any sort of arguments regarding privacy invasion in private space. Consent/permission seeking was considered important. They also expressed that approval of consent was equivalently important. Bystanders were expected to respect privacy once they were informed about them wanting it.

*"I expect people to understand that unless I am giving them a thumbs up to come and intrude my space whenever they feel like, they should not."*

*"I expect people to be generally considerate and not intrude my privacy if I ask them for some privacy. I do not expect people to know my limits of privacy. But I expect them to respect it once I let them know."*

Protecting oneself from privacy invasion in private space was considered as a personal responsibility. This thought was fueled by unfamiliar negative consequences that were products of digital era. These consequences were not anticipatory in nature. Participants wished that widely used applications should be responsible with protection of data. In comparison to public spaces, seven participants mentioned that they were less vigilant about invasion of privacy in spaces that they considered private.

*"I should be able to save myself from coming in the pics i don't want too"*

*"I would rather not be approached, filmed or bothered in these spaces. I think keeping people ought to keep to themselves and not intrude on what others do in these spaces. I don't want to be recorded or be compelled to share any information."*

*"Less cautious but still keeping an eye around"*

Two participants thought that unless their mental peace was disturbed in private space, invasion of privacy did not matter. They had developed the art of ignoring invasion of their privacy. A few participants thought that right to privacy in private spaces should be based on circumstances. Participants expected to receive auditory privacy from various corporations like Amazon. They felt that even if they gave consent, their privacy was being invaded. Protection of privacy in private space was considered unsuccessful as the presence of CCTV, facial recognition AI and location tracking through smartphone was always present. Receiving complete privacy was considered an illusion.

*"I've developed the art of ignoring things"*

*"I find it hard in the times of surveillance to protect oneself from such intrusion by the Government, Corporations etc. Even though the questions above assume visual intrusion and privacy, there is also auditory privacy. Amazon Echo uses our voices to train their ML models on servers located on unknown indigenous lands. So I don't know how to regulate and make policies what is private and what is public anymore?"*

**5.2.2 Privacy awareness**

Awareness is based on an individual's attention, perception and cognition of physical as well as non-physical objects [61]. It was found that participants were aware of unreliable internet platforms, illegal and unethical hacking, identity thefts, dark web, spoof attacks, information stealing and various other ways how organizations use data to lay schemes, memberships and deals. They were also aware of how Google history could be misused. A few participants expressed about technical problems when any form of data was synced to cloud. Majority of them were aware of how social media could negatively impact their life. They were cognizant of ways how privacy in various devices could be adjusted to suit their best needs. They mentioned being aware of how pictures over Facebook and Instagram were and could be misused. Participants additionally mentioned they were aware of how information available online could be used to commit crimes in their immediate physical environment hence, establishing physical boundaries was important. They were mindful of several drawbacks of surveillance and surreptitious recording, but also acknowledged the importance of CCTV cameras for vigilance. Participants who lived alone were more cautious and observant about surreptitious recording and applications in their smartphones that required them to give permission for any kind of recording.

*"Cover my webcam and have great security on my laptop. Avoid dodgy websites and links. Be extra careful on social media"*

*"Stay of the dark web and spoof attacks"*

*"make social media profiles private, installation of cctv cameras to keep vigilance etc"*

### 5.2.3 Consent or permission seeking

Consent means to give approval/ compliance in or approval of what is done or proposed by another [44]. Participants had a core belief that seeking consent and permission was a moral responsibility (in public space) and basic mannerism (in both public and private space). Participants desired either seeking permission or informing while clicking pictures in public spaces if there were any chances of invading bystanders' privacy. They also expressed that any bystander should seek permission and wait for its approval in private space.

*"I don't wish anyone to invade my private space without my consent"*

*"I expect people to understand that unless I am giving them a thumbs up to come and intrude my space whenever they feel like, they should not"*

### 5.2.4 Dilemma and confusion

Confusion is a situation in which a difficult choice has to be made between two different things one could do; uncertainty about what is happening, intended, or required [73]. Participants had confusion if technologies like facial recognition AI, Amazon Echo, CCTV surveillance was of help or invading privacy. Reliability of these platforms was questioned. They felt that consent and permission were not always honestly taken. Fear cultivated around the core belief that rapid spread of false information was greatly present. Participants also suspected about hidden camera and mic recordings that they were not aware about.

*"Even though I escape the cameras of people I can't escape the CCTV cameras of Shopping malls that are installed without my permission and are meant for my security in case of theft. I find it as a misnomer"*

*"The only concerns are hidden cameras or mics, otherwise no issues."*

*"Not to be too loud while discussing something important and maintain distance"*

### 5.2.5 Behavior or actions

Actions are self-initiated sequence of movements, usually with respect to some goal. It may consist of an integrated set of component behaviors as opposed to a single response [74]. Actions taken by participants to protect invasion of privacy in both public and private spaces were:

Participants remarked that if they felt that their privacy was violated then they would politely ask the violator to not invade their privacy. They avoided instances in public places where they thought that their privacy would be violated. They avoided sharing honest and complete information about themselves in restaurants and social media. Participants also ensured that they did not share any kind of personal information over Internet unless it was a reliable platform. Micro steps like not revealing full name on social media networks, not sharing pictures or videos of private life was repeatedly prevalent. Social media accounts were kept private as well. A few participants ensured that they have read privacy policies of various application. Four participants avoided using online banking applications frequently in public spaces mostly. In office space, participants used password protection to secure their smartphones, laptops and computer. Nine participants avoided using and holding an account on social media completely. They always turned their Bluetooth off. All kind of suspicious website and links were avoided, webcams were covered and disabled. Data security software was installed on their laptop. The participants refused to sign any schemes and membership deals. They used secure communication apps, turned off all Google history like location, search, watch, calendar and restricted or deleted all social media that linked them to oneself. Privacy in private space was maintained by taking precautions like shutting doors, windows, checking surroundings, installing secured lock systems and CCTV cameras.

*" do not mention complete details on social media; make sure I move out of frame if some one is taking a picture in public; making sure to read privacy of the application I use, ensuring that there are no cameras fixed inside public restrooms, hotel rooms before I use them, and so on.*

*" I make sure to lock my laptop in office so that my crucial data isn't available to all, also I have protected screen lock in my smartphone..."*

<u>Participants undertook the following steps to protect privacy of themselves in public spaces-</u>

They established boundaries to prevent encroachment of space, gadgets were secured with lock to keep information safe. Forty-three participants revealed that they were always very cautious, vigilant and alert when they were in public. They maintained distance, avoided crowded places, maintained low profile of oneself and maintained discretion of oneself by wearing masks and used earphones. One participant mentioned that they put on a poker face so that people mistook them for being unapproachable. Participants reported that they did not open applications which required them entering passwords and pins in public space.

*"Keep my bluetooth off and probably avoid crowded spaces. Trust my intuition."*

*"Less cautious but still keeping an eye around."*

*"Limit yourself giving too much information over the internet"*

Figure 3 below represent the themes discovered in this study. This figure shows how user perception is formed in both public and private spaces. Attitude, beliefs, ideas towards privacy, awareness about various privacy risks and regulations, consent seeking and various dilemmas influenced actions and behaviour of individuals. Interactions among all the themes generated experiences which influenced to form user perception towards privacy.

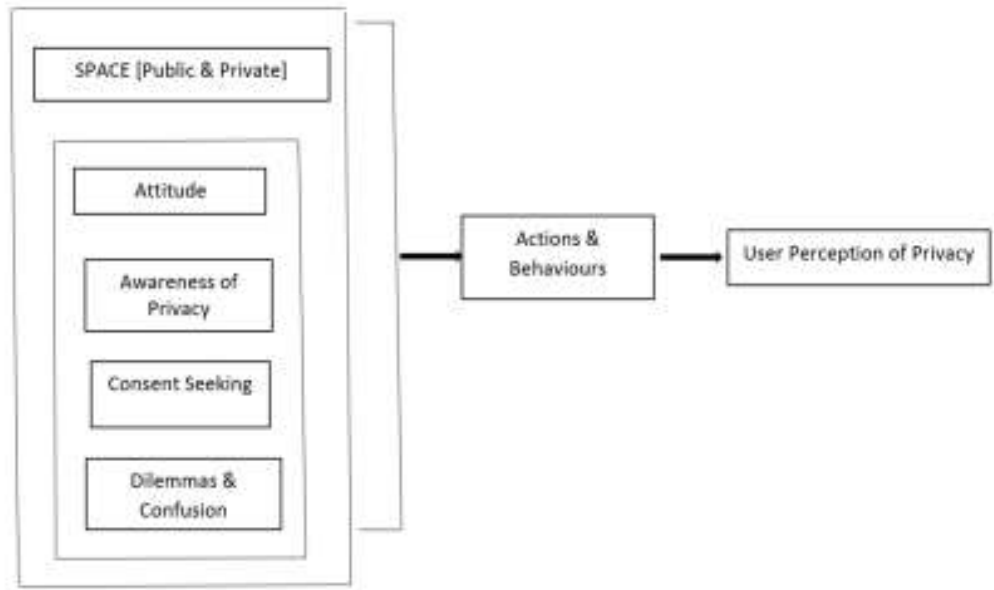

Figure 3: Represents how user perception is formed. It is the interplay of attitude, awareness, consent, confusion and actions that influences in the formation of user perception.

The next section would explain the themes discovered in this section in a detailed way. These themes are contextualized with relevant literature as much as possible.

## 6   DISCUSSION

Key themes identified in results section were: attitude towards privacy in public and private spaces, privacy awareness, consent seeking, dilemmas or confusions related to various technologies, impact of attitude and beliefs on individuals' actions regarding how to protect oneself from invasion of privacy in both public and private spaces. This section provides a rich description and interpretation of the experiences of participants, pointed out related literature supporting and contextualizing themes and drew connection among themes and subthemes by identifying similarities, differences and correlations. It would also help to understand the themes and their interconnections in a comprehensive way. This survey was conducted amidst pandemic; hence responses of participants might have been influenced because of the ongoing events. It was intended to collect a representative data from both female and male. All participants were educated and familiar with various ubiquitous technology especially smartphones and laptops. There were certain places that fell into the grey area (like religious places, gym) and could be intercepted as both private and public spaces. Participants respected privacy of bystanders in these ambiguous places. They demonstrated sensitivity in public space and declared that they would not capture pictures of women, children and people from lesbian, gay, bisexual, transgender and queer or questioning (LGBTQ) community.

### 6.1 Attitude towards privacy invasion

People held certain expectations regarding protecting themselves from invasion of privacy. Expectations are thoughts and beliefs that people hold regarding a phenomenon. In public spaces, some of the beliefs, thoughts and expectations of people regarding protecting oneself from invasion of privacy were-

All individual's actions are guided by interplay of thoughts and feelings. Protecting privacy of others in public space was considered a 'moral responsibility'. This was a core belief held by participants [21]. As reported by participants, they had a negative opinion towards photobombing, were uncomfortable when they were photographed without their knowledge, had a notion that pictures taken in public were mostly uploaded at various social media platforms and believed that every person had personal notions of boundaries. These thoughts guided their action to respect privacy of others in public, especially bystander's facial privacy. So, whenever they wanted a photo in public and there were people around, they waited for them to clear out and then engaged into clicking pictures. Also, they ensured to seek permission before clicking pictures in public from bystanders as they themselves were uncomfortable when consent was not seeked from them. These actions of seeking permission and waiting were guided by empathy and morality which is in accordance with the theory of moral development proposed by psychologist Lawrence Kohlberg. Participants were being respectful of others' privacy and seek consent because they were conforming and complying to social norms. [21].  These social norms were respecting privacy of others in public, maintenance of decency like not eavesdropping and starring. Only one participant reasoned the legal aspect for respecting privacy of people in public. This suggested that majority of the participants respected privacy of others considering more of it as fulfillment of social norms than legal rights.

Participants also reported that invading privacy of others in public could also be a mistake as there was a lot of ambiguity surrounding the consequences of invading privacy in public spaces. For instance, taking a picture in public place in India is not illegal as there's no such law which prohibits photography in public. But photographing somebody without their consent violates their right to privacy which is implicit in Art. 21 (Right to Life) of our constitution [15]. Photographing somebody is the first violation and then publishing it would lead to another violation. As mentioned in the results section, 73.33% of sample population avoided clicking pictures of women, children and any stranger in public spaces altogether.  Now, a lot of people may not be aware of the laws the way a trained professional would be. Lack of awareness could be because of information overloading and difficulty to understand complex legal terminology. Moreover, judgement of consequences in India was decided by police and on the basis of evidence. This process involved a number of ambiguities into what could be explicitly stated as legal and illegal. In addition to this, following up with legal procedures was a time and money straining process. [36]

In private spaces, consent seeking was considered important, approval of consent/permission was desired. In public space consent and approval of consent was considered as same part of an action but in private space approval of consent was viewed as a separate sub-action. Negative confrontations occurred in private spaces but it was not desirable by participants. They considered protecting oneself from invasion of privacy was a personal responsibility. This core belief stemmed out because the consequences of privacy invasion in digital era were viewed negatively, ambiguities existed and less control over information was prevalent. So, participants were fearful of consequences. Lack of knowledge about privacy risks and their solutions contributed to this core belief. According to Richmond Y. Wong, privacy by design was desired and expected from widely used applications [51]. According to Fuming Shih, intuitive trust and familiarity played an important role in willingness to trust various applications [51]. On the other hand, a few participants developed apathy towards invasion of privacy in private spaces. This is not a new phenomenon. Lack of control could generate feeling of numbness. Some individuals coped up and found solutions that best suited them on the other hand others felt loss of control. It was necessary to keep in mind that various psychological needs of human beings like need for autonomy, relatedness and competence should be fulfilled when one is using various technology irrespective of public or private space. [19, 60] Feeling lack of control over online data, malicious interface design, ambiguous boundaries in virtual world etc. could be contributing to feelings of apathy. [1]

### 6.2 Privacy awareness

Privacy awareness of an individual encompasses the attention, perception and cognition of whether others receive or have received personal information about him/her, his/her presence and activities, which personal information others receive or have received in detail, how these pieces of information are or may be processed and used, and what amount of information about the presence and activities of others might reach and/or interrupt the individual. [61]. The study revealed that participants were aware, respectful and sensitive about privacy of people in public spaces. This judgement is on the basis of face validity. In a study conducted by Amon, M. J [4] found that people without strong personal privacy preferences were especially likely to want to share photos, regardless of how the photo portrayed the subject. It also reported that the privacy condition led to a lack of concern with others' privacy. These findings suggested that developing interventions for reducing photo sharing and protecting the privacy of others is a multivariate problem. Participants in our study were aware about various types of privacy like privacy of behavior and action, privacy of communication, privacy of data and image. They were accustomed with the difference between violation and invasion of privacy and their diverse consequences (technical and non-technical). These were some of the concerns that were growing with emerging technologies that could be observed in both private and public spaces. It was not possible to gaze to what extent or degree was this awareness. Also, privacy invasion perceptions were negatively related to continuous use intentions [80]. It was observed that participants actually did guide their actions in accordance with their perceptions.

### 6.3 Consent or permission

Both seeking permission/consent and informing was acceptable in public spaces but only seeking permission was acceptable in private space. In public space, participants did not mention about waiting for approval after seeking permission. This implied that bystanders always considered to seek permission and waited for its approval. It was also observed that consent for all the intended course of action was necessary in public spaces. This thought escalated from the idea that any content could be uploaded in social media without actually seeking permission to upload it. On the other hand, in private space inability of bystanders to wait for response from the participant was a concern. In public space, when privacy was invaded, it was tolerated to some extent as participants held the belief that it could be because one is mistaken but in private spaces tolerance for invasion of privacy was reduced.

### 6.4 Dilemma and confusion

Dilemmas and confusion aroused as various emerging technologies came both with advantages and disadvantages. Implications of these technologies were heavily dependent on how they were used. Spreading awareness and learning about these emerging technologies could be helpful but not all are capable of it. As mentioned by some participants, various technologies have made people to be hyper-vigilant in this present era. Each of these technologies had capability to disrupt as well as contribute to our health and wellbeing (both physical and mental). Surreptitious recording was a big concern with participants as they used smartphones and laptops every day that were equipped to record not just visuals but also audio. CCTV and other forms of surveillance was a growing concern as participants were aware of various uses and consequences of the same. They knew about the pros and cons of AI and ML models but not having the skillset to prevent or control various technical damages could lead to technological anxiety. Awareness about these in easy simple ways could help to reduce technological anxiety. Participants mentioned that they used social media and read news that helped them to gain an understanding of various privacy related topics. These platforms could be effectively used to raise awareness about privacy risks, solutions and regulations.

**6.5 Behavior or action**

Thoughts and beliefs influence every individual's behaviour. Whenever conflicting situations related to privacy invasion aroused, participants were assertive and bystanders complied to the requests. Anticipated instances in public space where participants felt that there are chances of privacy risks, those situations were avoided mostly. This could be because certain human needs (like need for safety) [74] were not getting fulfilled. Participants broadly used two key ways to protect themselves from invasion of privacy namely avoiding privacy risk situations and discretion maintenance. The most preferred way was maintenance of distance and avoidance of crowded places. This could be because most participants engaged in flight response. When physical distance was maintained, it helped them feel safe and avoid feelings of uncomfortableness. Since violation of privacy could happen in multiple ways, participants wanted to feel safe as much as they could [17]. Participants claimed that they were more cautious and vigilant when in public. Feeling of unsafety could also lead to inadequate performance of intended actions. [18]. This feeling may possibly also instigate promotion of dishonest behaviour among people like sharing insincere information about oneself in various platforms like digital media and public spaces. This can have impact on various advertisement and promotional activities. When bystander's privacy is violated content of photo can be misused. Participants mentioned that any kind of transaction activity and applications that required entering of passwords and pins were avoided in public space. It was perceived as risky because actions could be captured or recorded. This thought was contributing to usage of mobile payment in public spaces in less frequency. Mostly data security apps were used in both public and private spaces.

It was found that in public spaces, participants respected privacy of people around as they believed that it was their moral responsibility. In private space it was felt that protecting oneself from invasion of privacy was a personal responsibility. In both the spaces, seeking consent by bystanders was desired. Actions of participants was in line with their thoughts. This research would contribute to understand user's perspective of privacy in public and private spaces. Negative attitude towards surveillance and other privacy concerns can be addressed by various organizations by taking into account privacy by design. These findings could contribute to design socio technical solutions, positive and persuasive technologies and behavioural interventions. The following section highlights some of the interesting findings in the upcoming section. This section would also briskly discuss the results-

**6.6 Summary**

The following bullet points summarized the paper and study:

- Protecting oneself from invasion of privacy was considered as a personal responsibility. It meant that participants wanted to fulfil their need for autonomy while protecting themselves from privacy invasion. This should be considered as a design principle while designing immersive technologies. Participants trusted themselves and their intuition when it came to protecting oneself and preventing invasion of privacy. Figure 4 below is a pie diagram that represented various causes cited by participants as to why they should protect themself from invasion of privacy and considered it personal responsibility.

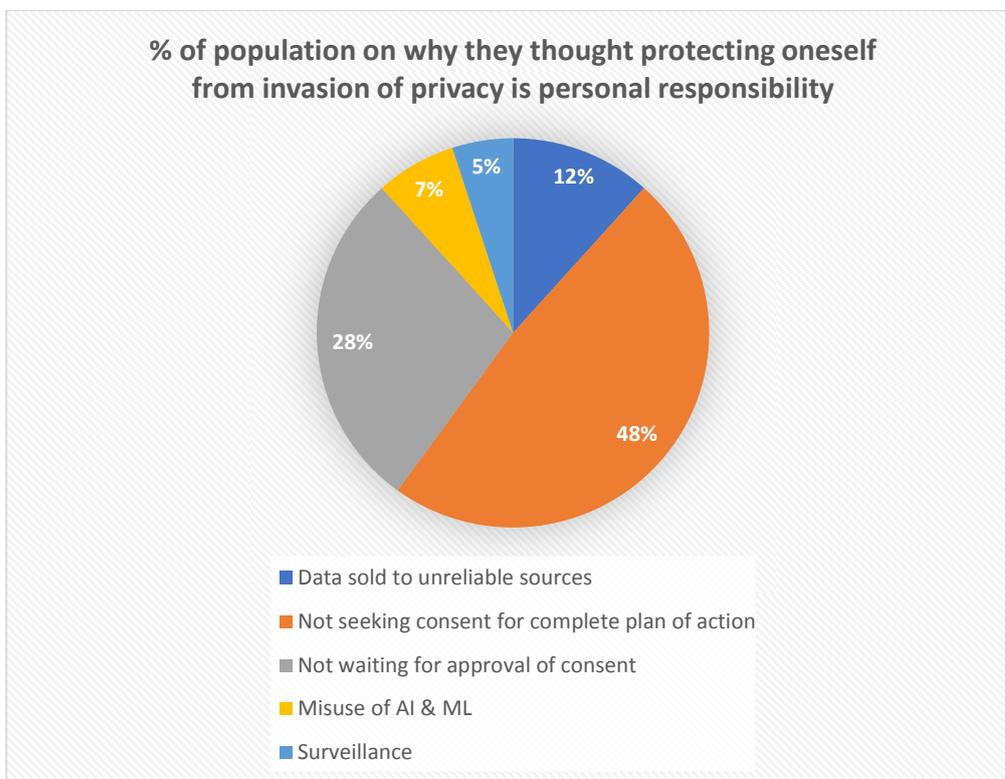

Figure 4: Pie diagram showing percentage of population as to why they thought protecting oneself from invasion of privacy is personal responsibility

- Social media was considered a reliable tool to keep oneself aware of various privacy regulations. Increasing awareness of various privacy regulations on social media could build trust and also influence attitude of participants.
- Participants respected privacy of others in public because they held a core belief that it was a moral responsibility. It was in line with moral development theory.
- Participants trusted their close friends and professional photographers the most to click their pictures irrespective of space. Colleagues were preferred third amongst the choices provided. Attitude can be best influenced by targeting these key players. This would be useful for advertising, sales and marketing professionals to decide target population criteria.

Figure 5 below represented percentage of population who were comfortable getting clicked by various people. Least number of participants were comfortable getting clicked by media personals. On the other hand, preference for getting clicked by professional photographer and colleagues had a close margin. 31% of participants were comfortable getting clicked by friends.

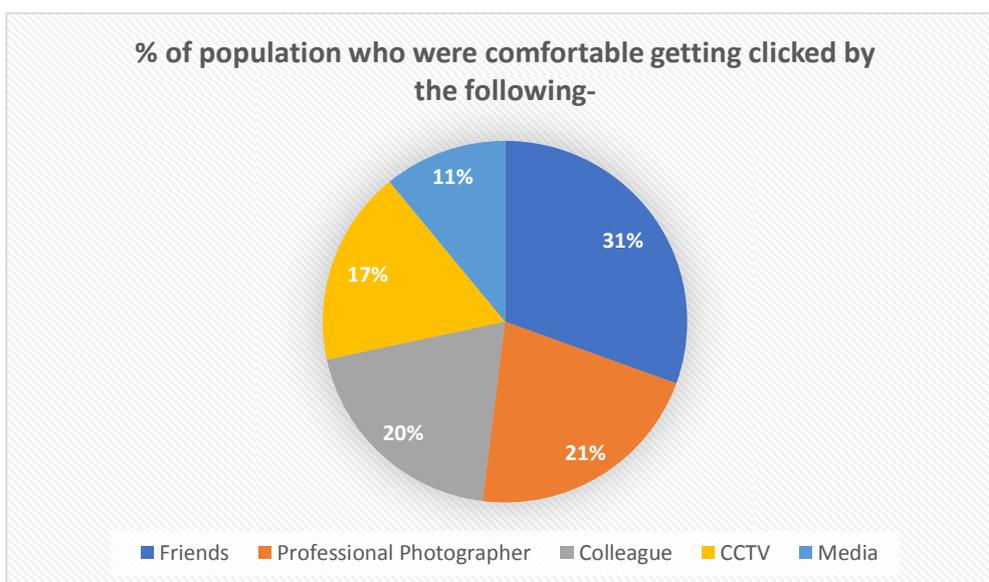

Figure 5: Pie diagram showing percentage of population who are comfortable getting clicked by various sources

- Actions taken to protect oneself from invasion of privacy in digital space were: having minimal presence online, maintain a pseudonymous self, holding no social media accounts. These were the spectrum of behavior which are representative of fear/ negative consequences surrounding invasion of privacy.
- Attitudes that influenced fear of invasion of informational privacy were: data being sold to unreliable and untrustworthy sites; not seeking consent for entire plan of action. For example- seeking consent for clicking picture but not for uploading; seeking consent but not considering approval of consent; being watched and monitored through webcams and hacking; misuse of machine learning (ML) models and artificial intelligence like facial recognition; invasion of auditory privacy through ML models and miscellaneous video recording (surveillance). Transparent awareness of various ethics followed concerning to the most commonly occurring fear/ concerns of participants can be promoted over the platforms that they choose to keep themselves aware of various privacy regulations.
- The need for autonomy to protect oneself from invasion of privacy was because participants were less aware about privacy risks. This lack of awareness pushed participants to engage in behavior like avoiding public places altogether and their presence in social media.

It was found that despite the existence and importance laid on privacy risks, participants were not aware of the implications in full potential. A number of participants were not aware of how various information is collected and used. By understanding opinions and perspectives from general population, stakeholders could enhance privacy in their products. It would also help to better recognize the cognitive model of user's privacy needs. There have been privacy concerns since decades but currently concerns related to implications and impact of various emerging technologies are intensely growing. It is important to acknowledgment the fluid nature of privacy with respect to spaces and accommodate them in various applications.

There are different types of privacy. They may be applicable for both public and private spaces but context changes. Group privacy was desired and eavesdropping was unsought in public spaces. There are no technical solutions available for these concerns yet. Also, there were a number of solutions to address bystanders' privacy concern but none of the participants mentioned about using any technical solutions to protect themselves from invasion of facial privacy in public space. A number of participants did use technical solutions to protect their privacy in private space.

**7 CONCLUSIONS**

In the present age of overwhelming number of ubiquitous devices with in-built cameras, understanding and preserving privacy of users and bystanders are of paramount importance. In particular, many vulnerable users like people with disabilities, neo-literate persons and often their able-bodied counterparts often do not realize their privacy is compromised. In this context, the present study is designed to understand users' perspective of privacy. Considering the ongoing Covid-a9 pandemic, we used an online survey and qualitative research method. Key themes found in this study were attitude towards privacy in public and private space, privacy awareness, consent and permission seeking, dilemmas regarding $S^3$ and actions or behavior. Participants expressed that the Government should come up with rules that would hold the perpetrators accountable who invaded privacy of others in public space with regard to facial privacy. Explicit regulations and policies regarding protection of privacy was required as not many were aware of it. Prevalence of confusion regarding what actions to be taken post violation of privacy should be well circulated. These findings can help to develop various ubiquitous technologies, positive and persuasive technologies. Seeking permission was a widely practiced behavior as individuals wanted to avoid discomfort and negative consequences. It was influenced by the core belief of 'respecting privacy is a moral responsibility'. This study was not representative of gender and data from more sample could be collected. Moreover, this study primarily focused on the opinions and thoughts of the educated. Despite trying our best, the online survey might not have been able to address the dynamic nature of privacy invasion. Future work could be exploring and understanding implications of privacy invasion on mental health leading to apathy.

# A    APPENDICES

**Survey Questionnaire**

1. Name? ______________________

2. Gender?

a) Male

b) Female

c) Other

3. Age (in years) *

18-24

25-34

35-44

Above 44

4. Native Place *

[ ]

Your answer

5. Place of current residence *

[ ]

Your answer

6. Mother Tongue *

[ ]

Your answer

7. Ethnicity *

- African American
- White
- Asian
- American Indian or Alaska Native
- Native Hawaiian or Other Pacific Islander
- Hispanic or Latino

8. Highest educational qualification *
- High School ( from class 9 to 12)
- Under Graduation
- Post Graduation
- Other:

9.How good are you with technology? *

- Beginner
- Good
- Tech Savvy

9. Name a gadget that you use most often on a daily basis. *
- Smartphone
- Laptop
- Computer
- Television
- Gaming devices
- Non smart mobile phones

10. Name any other gadgets you use on a daily basis. *
- Smartphone
- Gaming devices
- Laptop
- Computer
- Other:

11. How often do you take pictures/videos of oneself and others? *
- Frequently
- Once in a day
- Once in a week
- Once in 15 days
- Other:

12. What public places do you visit the most? (Multiple select question) *
- Public transport
- Restaurants
- Gyms and Fitness centers
- Parks
- Shopping Malls

Other:

13. How often do you visit public places (E.g.: Shopping malls, Public transport etc)? *
- Daily
- Once in two days
- Weekly
- Monthly

14. Do you take pictures and videos in public places? *
- Yes
- No

15. How often do you attend video calls in public? *
- Frequently
- Once in a day
- Once in a week
- Once in a month
- Very rarely

16. How often have you found stranger/s in the pictures that you clicked in public places? *
- 1 to 3 times
- 4 to 5 times
- More than 5 times
- Other:

17. Are you comfortable being photographed by (can select multiple options)- *

- Stranger
- Media person
- Professional photographer
- Friend
- Colleague
- CCTV Camera

18. Are you comfortable with unexpected appearance of oneself in the camera's field of view in a picture taken by(can select multiple options) - *
- Media
- Stranger
- Professional Photographer
- Friend
- Colleague
- CCTV
- Other:

19. As a person clicking pictures/videos of oneself and others, would you respect the privacy preferences of other people around you? *
- Yes
- No
- Maybe

20. Please support your answer with rationale for the above question! *

Your answer

21. As a person clicking pictures/taking videos would you avoid taking pictures/videos of- *
- Children
- Women
- Other:

22. Name a few public spaces where you would like to avoid taking pictures. *

Your answer

23. Name a few places that you consider it as private space? *

Your answer

24. What are your expectations regarding protecting oneself from invasion of privacy in your private spaces? *

Invasion of privacy is the unjustifiable intrusion into the personal life of another without consent

Your answer

25. What are your expectations regarding protecting oneself from invasion of privacy in public spaces? *

Invasion of privacy is the unjustifiable intrusion into the personal life of another without consent

Your answer

26. Can you please mention a few measures that you personally take to protect your own privacy in public spaces? *

Your answer

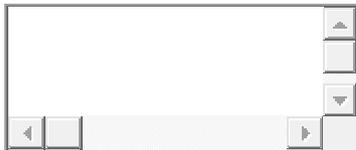

27. Can you please mention a few measures that you personally take to protect your own privacy in private spaces? *

Your answer

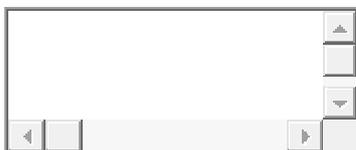

28. Please name a few sources through which you stay updated about various privacy regulations? *